\documentclass{SciPost}

\hypersetup{
    colorlinks,
    linkcolor={red!50!black},
    citecolor={blue!50!black},
    urlcolor={blue!80!black}
}
\usepackage[bitstream-charter]{mathdesign}
\usepackage{amsmath}

\usepackage{graphicx}
\usepackage{subcaption}  % modern an
\usepackage{dsfont}
\usepackage{comment}
\usepackage{fancyhdr}
\usepackage{slashed}
\usepackage{float}
\newcommand{\beq}{\begin{equation}}
	\newcommand{\eeq}{\end{equation}}

\def\nn{\nonumber}

\def\be{\begin{eqnarray}}
	\def\ee{\end{eqnarray}}
\newcommand{\bea}{\begin{eqnarray}}
	\newcommand{\eea}{\end{eqnarray}}

\DeclareSymbolFont{usualmathcal}{OMS}{cmsy}{m}{n}
\DeclareSymbolFontAlphabet{\mathcal}{usualmathcal}

\fancypagestyle{SPstyle}{
\fancyhf{}
\lhead{\colorbox{scipostblue}{\bf \color{white} ~SciPost Physics }}
\rhead{{\bf \color{scipostdeepblue} ~Submission }}

\fancyfoot[C]{\textbf{\thepage}}
}

\begin{document}

\pagestyle{SPstyle}

\begin{center}{\Large \textbf{\color{scipostdeepblue}{
%%%%%%%%%% TODO: Write your article's title here
Fermi Liquid Fixed Point Deformations due to Codimension Two Defects\\
}}}\end{center}

\begin{center}\textbf{
% Use (full) first name (+ middle name initials) + surname format.
% Separate subsequent authors by a comma, omit comma and use "and" for the last author.
% Mark the corresponding author(s) with a superscript symbol in this order
% \star, \dagger, \ddagger, \circ, \S, \P, \parallel, ...
Jin-Yun Lin\textsuperscript{1$\star$},
Ira Z. Rothstein\textsuperscript{1$\dagger$}
}\end{center}

\begin{center}
{\bf 1} Department of Physics, Carnegie Mellon University, Pittsburgh, USA

%%%%%%%%%% END TODO: AFFILIATIONS
% Provide email address of corresponding author(s)
$\star$ \href{mailto:email1}{\small jinyunl@andrew.cmu.edu}\,,\quad
$\dagger$ \href{mailto:email2}{\small izr@andrew.cmu.edu}
\end{center}

\section*{\color{scipostdeepblue}{Abstract}}
\textbf{\boldmath{%
We show that codimension-two defects in Fermi liquids
deform the renormalization group flow via a marginally relevant coupling. The mechanism for generating the flow is distinct
from the case of the Kondo problem (codimension-three defects)
in that the effective particle-hole asymmetry that leads to the log running is due to the spatial anisotropy generated by the defect.
The mechanism for the log generation  has a simple geometric explanation which shows that hole fluctuations are suppressed as the incoming momentum 
is taken to be along the direction of the defect.
The RG flow time is shown to scale with the length 
of the defect. We also show that the dislon, the Goldstone mode localized to the defect, couples in a non-derivative fashion to the bulk fermions
and  becomes relevant above the dislons' Debye frequency which depends upon the defect tension.
}}

\vspace{\baselineskip}

%%%%%%%%%% BLOCK: Copyright information
% This block will be filled during the proof stage, and finilized just before publication.
% It exists here only as a placeholder, and should not be modified by authors.
% \noindent\textcolor{white!90!black}{%
% \fbox{\parbox{0.975\linewidth}{%
% \textcolor{white!40!black}{\begin{tabular}{lr}%
%   \begin{minipage}{0.6\textwidth}%
%     {\small Copyright attribution to authors. \newline
%     This work is a submission to SciPost Physics. \newline
%     License information to appear upon publication. \newline
%     Publication information to appear upon publication.}
%   \end{minipage} & \begin{minipage}{0.4\textwidth}
%     {\small Received Date \newline Accepted Date \newline Published Date}%
%   \end{minipage}
% \end{tabular}}
% }}
% }
%%%%%%%%%% BLOCK: Copyright information

% For convenience during refereeing we turn on line numbers:
%\linenumbers

\vspace{10pt}
\noindent\rule{\textwidth}{1pt}
\tableofcontents
\noindent\rule{\textwidth}{1pt}
\vspace{10pt}

% Write your article contents here, starting from first \section.
% An example structure is given below.

\section{Introduction}

\begin{figure}[htb]
  \centering
  \begin{subfigure}[t]{0.48\columnwidth}
    \centering
    \includegraphics[width=0.6\linewidth]{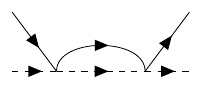}
    \caption{Particle Channel}
    \label{fig:diagA}
  \end{subfigure}
  \hfill
  \begin{subfigure}[t]{0.48\columnwidth}
    \centering
    \includegraphics[width=0.6\linewidth]{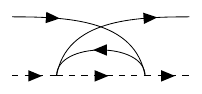}
    \caption{Hole Channel. }
    \label{fig:diagB}
  \end{subfigure}
  \caption{One-loop correction to the Kondo coupling.}
  \label{fig:two_diagrams_column}
\end{figure}

The Kondo problem is a paradigm of the renormalization
group (RG) in many-body physics in which a magnetic impurity (point defect) leads to a marginally relevant coupling with electrons
in a Fermi liquid. As the temperature is decreased the coupling
grows, leading to the famed "resistivity minimum" \cite{10.1143/PTP.32.37} which arises
as a consequence of the RG-induced strong coupling\cite{RevModPhys.47.773, PWAnderson_1970, Hewson1993Book}. 
If one just considered scattering off a delta-function potential, 
no log would arise in a three-dimensional Fermi liquid with a smooth Fermi surface\cite{Abrikosov1963},  for which the necessary condition is the existence of a dynamical degree of freedom on the point defect. However, this condition is not sufficient. 
% For instance, one
% could imagine that the point defect has some low-lying excited state that generates dynamics in the impurity. 
Within dynamical impurity models, the fate of the logarithmic correction is organized by the so-called commutative/non-commutative classification, as reviewed in \cite{cox1997exotickondoeffectsmetals}.  
The Kondo problem is a prototypical example of the non-commutative class, where the cancellation of the particle and hole channel of the one-loop correction (see Fig. \ref{fig:two_diagrams_column}) is prevented by having distinct spin structures for the particle and hole contributions, i.e. the diagram generates
$\log(\Lambda/E)([\sigma^a, \sigma^b]_{\alpha \beta}S^a S^b)$, where $S^a$ is the impurity spin.
In the commutative class, the impurity operators commute and the particle-hole cancellation survives. For example, see \cite{PhysRevB.26.1559}\footnote{Strictly speaking, logarithmic flow can also arise in commutative impurity models once electron–hole symmetry is broken \cite{PhysRevB.56.12947}. However, such effects require a dynamical impurity whose couplings reshape the local electronic environment and are not relevant for the static extended defect considered here.}.
Of course, one could also induce
particle hole asymmetry say by working near the bottom
of the band, but then we would be moving away from the class of Fermi liquids of interest in this paper. It is thus fair to say that RG flow due to impurity scattering is not generic.
\begin{figure}[h]
    \centering
\includegraphics[width=0.4\linewidth]
{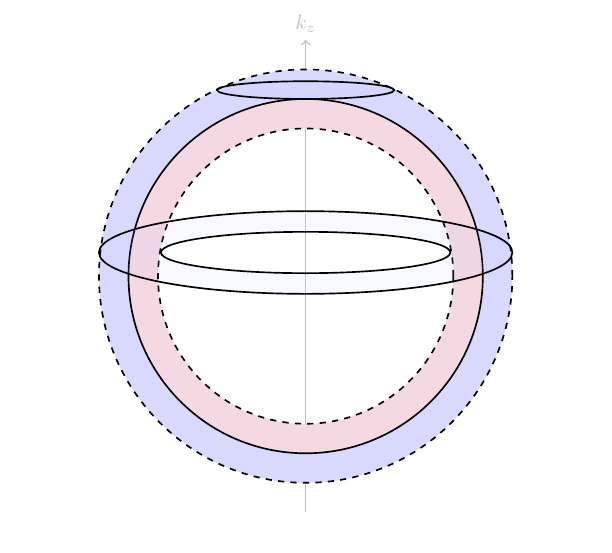}
\caption{Allowed intermediate states in one-loop scattering correction due to $z-$momentum conservation on the Fermi sphere. The solid sphere represents the FS, and the dashed outer/inner lines bound the particle/hole phase space in the EFT. We see that when the incoming momentum is along the defect (z-direction) the kinematics induce a particle-hole asymmetry since only particle states are available above the pole.}
\label{fig:FS_diagram}
\end{figure}

Here we point out that codimension-two
defects can induce a non-trivial RG flow without the need to introduce dynamical degrees of freedom on the defect. The origin of the logs is most easily understood in the context of the effective field theory (EFT) of Fermi liquids \cite{RevModPhys.66.129,polchinski1999}. If we consider expanding around the Fermi surface, then for generic momenta
the scalar coupling localized on the defect will not run 
as the particle and hole contributions will cancel each other out, as in the case of a point-like delta function potential. However, when the incoming momentum is along the preferred direction ($\hat z$) of the dislocation, the conservation of momentum forbids hole excitations as depicted in Fig. \ref{fig:FS_diagram}. The usual cancellation between particle and hole channels is thus kinematically obstructed and the generic coupling function for the localized
operators will run. We emphasize that the logarithmic flow produced here requires neither a dynamical impurity nor a non-commutative operator structure, and it preserves the particle–hole symmetry of the bulk Fermi liquid. The effect arises purely from the geometric constraint imposed by a codimension-two defect.

Complementary developments have highlighted the rich quantum structure associated with  dislocations and their effect on RG flows in non-Fermi liquids, such as emergent phenomena in Dirac materials due to dislocations\cite{Volovik_2015, PhysRevB.95.115410}, and a classification of renormalization group fixed points in connection to crystalline dislocations and disclinations\cite{barkeshli2025disclinationsdislocationsemanantflux}. The present work focuses on identifying a concrete mechanism by which extended defects can generate a nontrivial renormalization-group flow in an otherwise conventional three-dimensional Fermi liquid.

\emph{Conventions}: We work with the mostly plus metric $(-,+,+,+)$, Greek  (space-time) indices run $(0-3$) while Roman capital letters $I=1-3$
		correspond to the internal degrees of freedom (or, more precisely, the comoving coordinate system). 
		Euclidean spatial indices are represented by small Roman letters.
		Note that after symmetry breaking, we can no longer
		distinguish between the capital and small Roman indices.
		We will also reserve the index $a$ for directions transverse to the dislocation, i.e.
		$a=1,2$ while $i=1-3$.
		We work in units where $\hbar=c=1$.

\section{The EFT of Dislocations}

To formalize the deformation of the Fermi liquid RG flow
(we refer the reader to \cite{polchinski1999,RevModPhys.66.129} and \cite{Burgess_2020} for reviews of this EFT), we will
follow the recently developed EFT for dislocations {\cite{dislonPaper}} which extends the EFT of solids. Note that in solids there are two classes of  codimension-two defects, called dislocations
and disclinations. The mathematical distinction is that their associated
holonomies are translations and rotations respectively\cite{Kleinert2008, HirthLothe1982}.
In this paper we will only be concerned with dislocations since
they are much less energetic and more ubiquitous in solids. The direction of the translational holonomy (called the Burgers vector) leads to two classes of defects, screw and edge dislocations depending on whether the Burgers vector is parallel or transverse to the dislocation line. In this way, the classification only makes sense locally if the line is curved.

The defect breaks translational invariance in the transverse\footnote{At this point we will assume the defect is a straight line and will generalize later.} directions as well
as rotations away from its axis.\footnote{The lattice breaks translations down to some discrete subgroup.} However, a dislocation distinguishes itself from the fundamental string in that it breaks boosts along
the string so it does not share the reparameterization invariance of the fundamental string.  As such, there are three Goldstone bosons, two transverse\footnote{In the case of a straight edge dislocation, the so-called glide constraint prevents transverse excitation perpendicular to the Burgers vector.} and one longitudinal mode\footnote{These Goldstones have been called ''dislons" \cite{Li_2018}.}, the latter of which is absent in the fundamental string. These modes can in principle play an important role
in the RG systematics when we couple the dislocation to electrons.
As we will now show, in the low energy regime relevant to this work, the effects of these Goldstones will only
generate irrelevant operators on the world-sheet. This is perhaps
not terribly surprising, given that usually Goldstones decouple
in the IR, though there are exceptions when space-time symmetries are
broken \cite{Rothstein_2018, Watanabe_2014, PhysRevB.64.195109}.

We will define the embedding coordinate for the dislocation to be $X^\mu(\tau,\sigma)$, where
$\tau$ and $\sigma$ parameterize the world-sheet. Notice that we use covariant notation
here only for the purposes of simplifying formulae. We will eventually take the $c \rightarrow \infty $ limit.
As opposed to the fundamental string, the underlying lattice  breaks  boost invariance
along the string, so there will not be a full reparameterization invariance.
Despite this fact \cite{horn2015effectivestringtheoryvortex}, we can work in the static gauge
 where $X^0=t$ and $X^3=z$ at the cost of introducing a new degree of freedom corresponding to the
 longitudinal boson $\phi(t,z)$.
The action for the dislocation given by the most general functional of the fields 
consistent with all of the symmetries is \cite{dislonPaper}
\beq
S_{dis}=\int d\tau d\sigma  \sqrt{-g} F(B,\hat B).
\eeq
 The 
induced metric is $g_{\alpha \beta}= \eta_{\mu \nu}\partial_\alpha X^\mu \partial_\beta X^\nu$, $B= g_{\alpha \beta} \partial^\alpha \phi 
\partial^\beta \phi$ and $\hat B^\prime= h_{\alpha \beta} \partial^\alpha \phi 
\partial^\beta \phi$, where $h_{\alpha \beta}=G_{\mu \nu}\partial_\alpha X^\mu \partial_\beta X^\nu$.
$G_{\mu \nu}$ is the bulk metric in the solid that depends on the crystal structure. $F$ is some unknown
function of $B$ and $\hat B$. Note that we cannot Taylor expand in $B$ and $\hat B$ until  we  have chosen a
vacuum around which to expand, after which a  derivative expansion can be performed.
Expanding the action around a straight string $\langle \phi \rangle=z$, one can derive the action for the  dislons. 
To calculate the dispersion relation for the dislon modes, one must account for the fact
that they will mix with the bulk phonons which arises from the coupling of the string current
to the bulk phonons. The phonons also live inside the metric $G$, but the resulting couplings to the dislon
will only lead to non-linearities, which will not be of interest to us here. The result of the mixing leads to a non-analytic dispersion relation given in \cite{dislonPaper}.

\section{Power Counting }
\subsection{Review of Bulk Power Counting}

 The power counting parameter is defined as $\lambda\equiv E/E_F$  with the Fermi energy $E_F$. The quasiparticle energy is measured with respect to the chemical potential $\mu=E_F$:
\beq E(\vec k) = \epsilon(\vec{k})-\mu.\eeq
For low-energy excitations $(E,\vec k)$, while the energy scales towards zero, the corresponding momenta scale towards the Fermi surface with radius $k_F$. It is therefore useful to write a momentum close to the Fermi surface as 
\beq \label{eq:FSdecomp}
\vec{k} = \vec{k_{FS}}(\Omega)+k_r\hat{n}(\Omega)
\eeq
where \(\Omega=(\theta,\phi)\) labels a point on the Fermi surface and \(k_r\) measures the small displacement normal to it. For the spherical Fermi surface considered here, \(\vec k_{FS}(\Omega)=k_{F}\hat{r}(\Omega)\) and \(\hat{ n}=\hat{r}\).
The quasi-particle energy can be expanded as
\beq \label{eq:RGEnergyExpansion}E(\vec k) \approx \epsilon(\vec{k}_{FS})-\mu +k_r\hat{r}\cdot \nabla\epsilon|_{\vec{k}_{FS}} = v_Fk_r,
\eeq
so that the low energy scaling is
\begin{align}
E &\sim \lambda \nonumber \\
k_r&\sim \lambda \nonumber \\
\Omega  &\sim \lambda^0. \label{scale}
\end{align}
Thus, the residual momentum normal to the Fermi surface scales to zero, whereas the angular location on the Fermi surface does not scale.
The scaling of the kinematic variables fixes the scaling of the field via the standard kinetic term
\beq 
S = \int dt [d^3k]\psi^\dagger(t, \vec k)\Big(i\partial_t-\epsilon(\vec k)+\mu \Big)\psi(t, \vec k)\sim 1.
\eeq
 The low-energy propagating degrees of freedom scale as
$\psi(\vec{q},t)\mapsto \lambda^{-1/2}\psi (\vec{q},t)$, which has complementary scaling in real space $\psi(\vec{x},t)\mapsto \lambda^{1/2}\psi (\vec{x},t)$.
 
Naive power counting suggests that a generic four-fermion interaction 
\begin{align*}
\int dt \int \prod_{i=1}^4[d^3k _i]
\delta^3\left(\vec k_1+\vec k_2-\vec k_3-\vec k_4\right)V(\Omega_i)
\psi^\dagger\left(t, \vec k_1\right) \psi\left(t, \vec k_3\right) \psi^{\dagger}\left(t, \vec k_2\right) \psi\left(t, \vec k_4\right)
\end{align*}
is irrelevant; its marginality arises only because momentum conservation acquires enhanced scaling on special Fermi-surface configurations—forward scattering or back-to-back configurations (BCS). This can be demonstrated either by properly counting the momentum-conserving delta functions, or by using
the pattern of broken space-time symmetries \cite{Rothstein_2018}.
The key lesson is that the Fermi surface geometry creates nontrivial constraints on the solution of the momentum-conserving delta functions, hence enhancing the interaction on special configurations. Since we will need to adopt this argument when we consider the coupling to the dislocation, we review the textbook
argument for the enhanced scaling of the delta function\footnote{A detailed pedagogical discussion of how delta-function constraints must be solved within the EFT domain can be found in Chapter 15 of \cite{Burgess_2020}.}. 

The EFT only includes low-energy modes near the Fermi surface, thus, even if $\vec{p}_1$, $\vec{p}_2$, and $\vec{p}_3$ lie in the EFT, the naive solution to the delta constraints $\vec{p}_4 = \vec{p}_1+\vec{p}_2-\vec{p}_3$ is in general unacceptable, as it may place $\vec{p}_4$ outside the EFT domain. In order for the solution to remain in the EFT we must choose special kinematic configurations where two of the momenta in the delta function nearly cancel. This is the familiar BCS and forward-scattering geometries, where two incoming momenta nearly cancel in the former, while one incoming nearly cancels one outgoing in the latter. On these solutions, one transverse momentum integral is removed. In this sense, the delta function effectively contributes a factor of $\lambda^{-1}$, and the resulting operator is marginal.

\subsection{Coupling Dislocation to Fermi Liquid}
We now couple Fermi liquid operators to  the dislocation\footnote{The effect of dislocations on electronic band structure was recently studied in \cite{Fall_2025}.}. One can write down RPI-invariant terms on the world-sheet with bulk operators:
\beq S=\int d^2\sigma \sqrt{-g} \hspace{2pt}\mathcal{O}_{FL}(X^\mu(\sigma)),
\eeq
where $\mathcal{O}_{FL}$ is a set of operators composed of the fermions and their derivatives consistent with the non-linearly realized symmetries.
Expanding $X^\mu$ around the equilibrium position of the dislocation we have
\bea S_{int}
&\approx & \int dz dt  (\mathcal{O}_{FL}(t,z)+ (X^a \partial^a \mathcal{O}_{FL}(t,z))+....= S_{int}^{static}+S_{int}^X
\eea
$S_{int}^{static}$ denotes the first term in the expansion, which captures defect-localized interactions that do not excite dislons, in effect treating the dislocation as a static scattering center. $S_{int}^X$, on the other hand, captures all interactions between bulk operators and the dynamical dislons. 

In $S_{int}^X$, the dislon does not couple derivatively like a typical Goldstone, since the derivative here is transverse.
Moreover, the derivative itself will, in general, have a component tangential to the Fermi surface and, as such, will not scale with $\lambda$. One might then conclude that the expansion is not convergent. 
However, much as in the case of the phonon, the scaling of the dislon field will depend upon
whether we are operating above or below the dislocation's Debye frequency $\omega_D^d\equiv v_s^d k_F$\footnote{We are assuming $k_D \sim k_F$.}(where $v_s^d$ is the dislon sound speed), which we will take
to be of the order of the bulk phonon Debye frequency. When the RG scale is lowered below $\omega_D^d$, the dislon will be
a pure potential (off-shell) ``Coulombic" mode, and the scaling is read off from the spatial derivative term 
in the dislon action $X^a \sim \lambda^{1/2}$, whereas above $\omega_D^d$, the dislon field will have a radiative component and scale as $\lambda^{-1/2}$.
The action will then scale (below, above) the Debye scale as 
\bea
S_{int}^X=\int dt dz  (X^a(t,z) \partial^a (\partial^r \psi^\dagger \partial^r \psi))(t,z,x=y=0)) \sim (\lambda^{1/2}, \lambda^{-1/2}),
\eea
where we have taken the lowest n-body operator consistent with the symmetries for $\mathcal{O}_{FL}$, as operators with more fermion fields will be suppressed by powers of $\lambda$.
The derivatives $\partial^r$ stand for the fact that we can write down as many derivatives as we wish as long as they pick out the momentum along the Fermi surface.
It is very interesting to note that for temperatures above $\omega_D^d$, the electrons will be strongly coupled to the dislon degrees of freedom. We will be considering the case where the sound speed is sufficiently large that the low energy range
of the validity of the EFT will be large and the dislons will decouple from the electrons and the leading interaction is given by $S_{int}^{static}$—interactions in which the dislocation is effectively, infinitely stiff, in which case
the only coupling we will consider is
\beq
S_{int}=  \int dt [d^3p_1][d^3p_2]\\\delta(p_{1,z}-p_{2, z})\psi^\dagger(t,\vec p_1) \psi(t, \vec p_2)g(\theta_1,\phi_1, \theta_2, \phi_2),
\eeq
where the polar angles $\theta_i$ are measured with respect to the dislocation axis. One might wonder if the delta function constraining the longitudinal momenta has enhanced scaling that would render this operator relevant. Note that the constraint 
 $\delta(p_{in, z}-p_{out, z})$ can be solved directly without recourse to special geometry, unlike the case for the bulk four-point function. Geometrically, the intersection of the constant-z plane with the EFT states around the Fermi surface is an area that scales as $\lambda$ (either a circle or an annulus); in other words,  the resulting measure $d^2p_{out, \perp}$ after solving $p_{z, out}=p_{z, in}$ still scales as $\lambda$. Consequently, there is no delta function enhancement and the defect operator remains marginal. 
 
% We can see that in the limit $\lambda \rightarrow 0$, the delta function constrains $\theta_1 = \theta_2$. If we keep the width $\lambda = \frac{E}{E_F}$ finite but small, then a straightforward trigonometric calculation shows that $|\theta_1-\theta_2|\leq \sqrt{\frac{2E}{E_F}}$ for all possible values of $z$ momenta. Thus, without loss of generality, the polar arguments of the coupling function can be expanded around a reference angle $\theta_0$ such that $\theta_{1,2} = \theta_0 +\Delta\theta_{1,2}$, where $\Delta\theta_{1,2}$ scales as $\lambda^{1/2}$ in the RG. Thus we may expand the dependence on the polar coordinates in the coupling function:
Without loss of generality, the polar arguments of the coupling function can be expanded around a reference angle $\theta_0$ such that $\theta_{1,2} = \theta_0 +\Delta\theta_{1,2}$, where $\Delta\theta_{1,2}$ scales as $\lambda^{1/2}$ in the RG. Thus, after expanding the coupling function
\bea
    g(\theta_1,\phi_1, \theta_2, \phi_2)&= g(\theta_0+\Delta\theta_1,\phi_1, \theta_0+\Delta\theta_2, \phi_2)\nn\\
    &= g(\theta_0, \phi_1, \phi_2)+O(\lambda^{1/2}).
\eea
The leading component of the coupling function $g(\theta_1,\phi_1, \theta_2, \phi_2)$ is used in all subsequent calculations. 

Regarding azimuthal angles, further simplifications arise in the case of a screw dislocation 
$(\vec{b}\parallel\hat{z})$, where rotational symmetry about the dislocation axis ensures that 
$g$ depends only on the relative azimuthal angle, $\phi_1-\phi_2$. By contrast, for an edge dislocation $(\vec{b}\perp\hat{z})$ , 
rotational symmetry around the line is broken. The interaction is attractive when $g>0$.
\section{The Anisotropic Fermi Liquid EFT}

In order to calculate systematically in this case, we will need to modify the
EFT. In any EFT the UV scale, $k_F$ in this case, should only show up in Wilson coefficients,
much like a mass in heavy particle EFT. To accomplish this goal we scale
out $k_F$ from the theory such that derivatives will be small.
This is accomplished by a rephasing
\beq
\psi(\vec x,t)= \sum_{\Omega_i} e^{i \vec k_F (\Omega) \cdot \vec x}\psi_{\Omega_i}(\vec x,t).
\eeq
 After rescaling,  derivatives in the
$\Omega_i$ direction will scale like $\lambda^0$ whereas all other directions will scale
as $\lambda$. $\vec k_F (\Omega_i)$ is a vector on the Fermi surface with magnitude $k_F$. 
This is akin to what is done when decomposing the Fermi surface into patches \cite{polchinski1999, Lee_2018, PhysRevB.82.075127}.

But in our case, the lack of momentum conservation in the transverse directions
forces a change in the formulation of the theory.
Expanding the dispersion relation around the Fermi surface, we have
\beq
E(k)\approx v_f(\Omega) k_r+...
\eeq
where $k_r$ is the orthogonal distance to the Fermi surface and $v_F(\Omega)=v_F$, which is related to Euclidean components of the momentum $(k_x, k_y, k_z)$  
\beq
k_r=  \sqrt{k_{xy}^2+k_z^2 }-k_F.
\eeq
We adopt $\vec{k}_{xy} $ as shorthand for $(k_x,k_y)$ for brevity. 

\begin{figure}[h]
    \centering
    \includegraphics[width=0.8\textwidth]{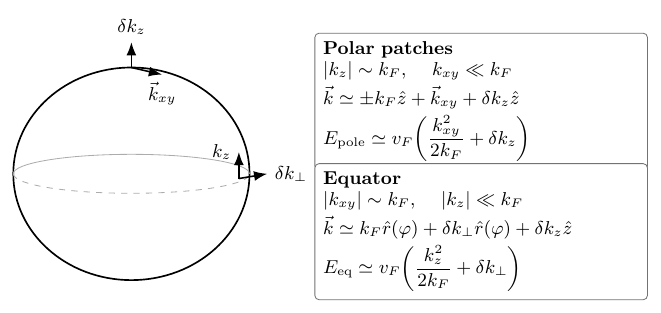}
    \caption{
    Schematic of the Fermi-surface momentum decomposition induced by the dislocation axis. The distinction between these regions is central to the anisotropic power counting used in the defect-coupled Fermi liquid EFT.
    }
    \label{fig:FS_schematic}
\end{figure}

The lack of momentum conservation will force us to consider distinct cases: when $k_z \sim k_F$ and $k_{z} \ll k_F$. This distinction is important because scattering processes with different incoming $k_z$ will lead to different collective effects. In the first case, the polar angle $\theta\sim 0, \pi$, and we can expand $\vec{k} = (\vec{k}_{xy}, k_z)$ around the large component $\vec k_F(\theta=0,\pi) = \pm k_F\hat{z}$ as \beq\label{eq:decompositionPoles}
\vec{k} = \vec{k}_{xy}+(k_F(\theta)+\delta k_z)\hat z.\eeq In the polar region, the dispersion relation $E(k) = v_Fk_r$ is
\beq\label{eq:FSexpansionPoles}
E(k)\approx  v_F(\frac{k_{xy}^2}{2 k_F} +\delta k_z).
\eeq
While in the latter case, $\theta=\frac{\pi}{2}$ and we can instead expand $\vec k$ around the large x-y plane component $\vec k_F(\frac{\pi}{2}, \phi) = k_F(\cos\phi\hat{x}+\sin\phi\hat y) \equiv k_F \hat{\rho}( \phi)$ 
\beq \label{eq:decompositionEquator}\vec k= k_z\hat{z}+(k_F+\delta k_\perp)\hat{\rho}( \phi).\eeq 
This gives a different dependence on the momentum for the dispersion relations
\beq\label{eq:FSexpansionEquator}
E(k)\approx   v_F(\frac{k_z^2}{2 k_F} +\delta k_\perp). 
\eeq
As we scale towards the Fermi surface, $k_r\sim \lambda$. The dispersion in the two kinematic zones provides a systematic way to identify the interplay between the anisotropy introduced by the dislocation and the EFT scaling, which is summarized pictorially in Fig. \ref{fig:FS_schematic}.

% In the former case, we can expand the momentum as $\vec k=(k_F+\delta k_z)\hat z+\vec k_{xy}$, giving dispersion relation:
% \beq
% {\cal E}(k)\approx  v_f(\frac{k_{xy}^2}{2 k_F} +\delta k_z).
% \eeq
% As we scale towards the Fermi surface, $k_r\sim \lambda$, which implies
% that $(\frac{k_{xy}^2}{2 k_F} +\delta k_z)\sim \lambda$. While in the latter case, 
% $\vec k=\delta k_z\hat{z}+(\vec k_{xy F}+\delta \vec k_\perp)$, such that:
% \beq
% {\cal E}(k)\approx   v_f(\frac{\delta k_z^2}{2 k_F} +\delta k_\perp),
% \eeq
% where we have used the fact that $\vec k_{xy F}$ and $\delta \vec k_\perp$ are collinear and that the magnitude of $\vec k_{xy F}=k_F$.
In principle we could define two different fields around the two regions and expand the momenta about their respective large momenta. However, this would be an unnecessary formal step at this stage. Of course, we have a continuum of operators in between the poles and the equator. However, we are only considering the extremes for the purposes of illustration.
Each path corresponds to a given value of $\theta_0$ in the coupling function as we shall see in the next section. The running of the logs will be between the two scales
$k_F$ and $(k_F- k_z)\sim \cos \theta_0k_F $.

\section{RG Flow of the Marginal Dislocation Vertex}\label{RG_flow_section}

\subsection{The Infinite String Length}

\begin{figure}[H]
    \centering
\includegraphics[width=0.3\linewidth]{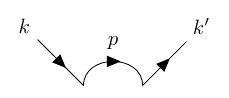}
    \caption{One-loop correction to the dislocation scattering vertex.}
    \label{fig:one-loop-defect}
\end{figure}

We will now show that the
coupling flows with an RG time interval set by the ratio $k_z/k_F$.  The one-loop correction $\Sigma$ to $g$ is given by the diagram in Fig. \ref{fig:one-loop-defect}. At each defect vertex, only \(z\)-momentum is conserved. The loop therefore includes an integral over the loop transverse momentum \(\vec p_{xy}\), as shown in Fig. \ref{fig:one-loop-defect}. The central point is that $\Sigma$ changes qualitatively due to the interplay between the local form of the  dispersion $E(k)$, introduced in the previous section, and the integral measure $\int_\Lambda [d^2p_{xy}]$, which selects all intermediate states that lie in the EFT cutoff while sharing the same $z-$momentum as the incoming mode.

Consider the case  where the incoming momentum $\vec k = (\vec
k_{xy}, k_z = k_F+\delta k_z)$ is near the North pole, which we call the "polar region".
The logarithmically divergent part of the  one-loop diagram, which we regulate by the EFT cutoff $\Lambda$, for these modes is given by
\bea
\label{one}
\Sigma (\theta_0=0, \phi_i-\phi_j)&=&(i)^3\int_\Lambda \frac{[d^2p_{xy}]}{\omega-E(\vec p_{xy}+(k_F+\delta k_z)\hat{z})} g(\theta_0,\phi_i-\phi_k) g(\theta_0,\phi_k-\phi_f) \nn \\&=& (i)^3 \int_\Lambda \frac{[d^2p_{xy}]}{\omega-v_f(\frac{p_{xy}^2}{2 k_F} +\delta k_z) } g(\theta_0=0,\phi_i-\phi_k) g(\theta_0=0,\phi_k-\phi_f) \nn \\
&=& i\frac{k_F}{v_F}\frac{I(\phi_i-\phi_j)}{(2\pi)^2} \log\left( \frac{\Lambda}{k_F( \delta k_z +\omega/v_f)}\right).
\eea
where the  angular integral is given by
\beq
I(\phi_i-\phi_j) =  (2 \pi)\sum_n   g_n^2(\theta_0=0)=\sum_n I_n.
\eeq
and we have decomposed the coupling in terms of spherical harmonics. Pictorially, $\int_\Lambda$ is the intersection between the constant $p_z-$plane and the low-energy states around the Fermi sphere, as highlighted in Fig. \ref{fig:FS_diagram}. The surviving large logarithm in $\Sigma(\theta_0 \in\{ 0, \pi\})$ is a reflection of the fact that $\int_\Lambda$ picks out predominantly particle states $E(\vec p)>0$ due to the kinematics. What we see here is that the kinematics is mimicking a particle-hole asymmetry as discussed in the introduction. The other pole leads to the same result. 

On the other hand, when the incoming momentum  $\vec k = (\vec{k}_{xy}, k_z)$ lies in the equatorial region $(k_z\ll k_F)$,  the loop momentum \beq \vec{p}=\vec p_{xy}+k_z\hat{z}=(k_F+\delta p_\perp)\hat{\rho}(\phi)+k_z\hat z\eeq  can be expanded in the manner of Eq. \eqref{eq:decompositionEquator}. In contrast to the polar case, the one-loop logarithm $\Sigma$ now cancels because the EFT shell is symmetric in \(\delta p_\perp\).
\bea
\label{two}
\Sigma (\theta=\frac{\pi}{2}, \phi_i-\phi_j)&=& (i)^3\int_\Lambda \frac{[d^2p_{xy}]}{\omega-E(\vec p_{xy}+ k_z\hat{z})} g(\theta_0=\frac{\pi}{2},\phi_i-\phi_k) g(\theta_0=\frac{\pi}{2},\phi_k-\phi_f) \nn \\
&=&(i)^3\int_\Lambda  \frac{k_Fd\delta p_\perp}{(2\pi)^2}\frac{1}{\omega-v_F(\frac{k_z^2}{2k_F}+\delta p_\perp)}I(\phi_i-\phi_f)\nn \\
&=& 0.
\eea
We expanded the measure $d^2p_{xy}$ in $2D$ polar coordinates $(p_r, \phi)$ and took the EFT expansion $p_r = k_F+\delta p_\perp$ to extract the leading marginal contribution. The lack of the log is a consequence of the fact that $\delta p_\perp$ is symmetrically integrated $-\Lambda <v_F\delta p_{\perp}<\Lambda$ in the equatorial region, as illustrated in Fig. \ref{fig:FS_diagram}.

%The kinematics are such 
%that when the incoming momentum is along the direct transverse to the defect the fluctuations get contributions for both particles 
%and holes in a symmetric fashion.

\begin{comment}
As reviewed above, this kinematic
constraint implies that the bare defect vertex is marginal at tree
level.  We now show that one-loop quantum corrections generate a logarithmic
flow of the vertex, and identify the
infrared cutoff set by the leakiness scale that weakly relaxes $k_z$.
The one-loop diagram (Fig. \ref{fig:one-loop-defect}) is given by the integral:
\beq Fig.~3 = \int \frac{d^3\mathbf{p}}{(2\pi)^3} 2\pi \delta(p_z-k_z)g(\mathbf{k}, \mathbf{p})g(\mathbf{p}, \mathbf{k}')G(\omega, \mathbf{p})\eeq
\end{comment}

\begin{comment}
The detailed calculation is shown in Appendix \ref{appendix:A}. Just as in the Kondo one-loop correction (fig. \ref{fig:two_diagrams_column}), the loop momentum describes a quasi-particle when $|p|\geq k_F$ and quasi-hole when $|p|\leq k_F$. 
\beq
\frac{k_F}{2\pi v_F}\mathcal{K}(k_z, \theta_0, \phi_{in}, \phi_{out})\Big(\log\Big(\frac{|\omega-\epsilon_{min}(k_z)|}{|\omega-v_F\Lambda|})+i\pi\text{sgn}(\tilde{\omega})\Big)
\eeq
where $\tilde{\omega}=\omega-\epsilon_{min}(k_z)\Theta(|k_z|\geq k_F-\Lambda)$,
\beq \mathcal{K}(k_z, \theta_0, \phi_{in}, \phi_{out})= \int \frac{d\phi}{2\pi} g(k_z, \theta_0,  \phi_{in}, \phi)g(k'_z,\theta_0, \phi, \phi_{out})\eeq
and $\epsilon_{min}(k_z)$ stands for the minimum energy in the EFT (measured with respect to the FS) for all modes $(\omega, \mathbf{p}) $ satisfying $p_z = k_z$. 
\beq \label{eq:energy_minimum_def}\epsilon_{min}(k_z) = \begin{cases}
v_F(|k_z|-k_F)&  |k_z|\geq k_F-\Lambda\\
-\Lambda & |k_z|\leq k_F-\Lambda
\end{cases}\eeq
so that near the caps $\epsilon_{\min}(k_z)$ vanishes linearly as $|k_z|\to k_F$.
\end{comment}

This result makes the mechanism for logarithm generation explicit.  For
$|k_z|\ll k_F$, the particle and hole contributions within the RG shell are
symmetric, and the leading logarithmic term cancels. Only when
$|k_z|\sim k_F$, i.e., when the incoming momentum is closely aligned with the defect axis, does the logarithm survive with a nontrivial energy dependence. The resultant beta function for the coupling function with $|k_z|\sim k_F$ is given by
\beq \label{eq:clean_beta_function}
 \frac{d g_n(\theta_0=0)}{d\ln \Lambda} = -g_n(\theta_0=0)^2 \frac{k_F}{(2\pi) v_f}.
\eeq
Thus, we see that for attractive UV data $g_n(\theta_0, \Lambda) > 0 $, the interaction is marginally relevant in the IR; for repulsive UV data $g_n(\theta_0, \Lambda) < 0 $, the interaction is marginally irrelevant.  

% The structure of the one-loop divergence in our problem is similar to the mechanism identified in field-theoretic analysis of branes and orbifolds, where bulk-brane interactions lead to renormalization of brane-localized couplings \cite{Goldberger_2001}. In the case discussed here, the classification of the divergence based on the codimension of the defect is distinct from \cite{Goldberger_2001}, because the propagator is that of Fermi liquid quasiparticles rather than a bulk relativistic field with homogeneous scaling. The scaling variable is the displacement from the Fermi surface, not any individual Cartesian momentum component. Consequently, the Cartesian transverse momenta integrated over because of the defect are not necessarily the same as the scaling variables of the Fermi liquid EFT. 
%In the Goldberger–Wise setup \cite{Goldberger_2001}, the brane interaction is localized by a delta function in the transverse directions, and therefore breaks transverse momentum conservation at the insertion. As a result, internal lines attached to the brane must be integrated over all transverse momenta. Evaluating the propagator at the brane position thus requires integrating over non-conserved transverse momenta up to a UV cutoff, which produces a short-distance divergence—power-law for generic codimension and logarithmic for codimension two.
The one-loop divergence resembles that of defect-localized couplings in relativistic brane theories \cite{Goldberger_2001}, where logarithmic scaling is naturally associated with codimension two. In a Fermi liquid, however, the scaling variable is the displacement from the Fermi surface rather than an individual Cartesian momentum component, so codimension alone does not determine the divergence. Moreover, logarithmic power counting is not sufficient for RG running: particle-hole cancellation must also be obstructed. This explains why codimension one does not generically exhibit the same flow by this static kinematic mechanism—the Fermi-surface regions with logarithmic scaling do not coincide with those where particle-hole cancellation fails. Codimension two is special for this mechanism because these two conditions align near the polar regions.

\subsection{Finite-Size Dislocations and RG Time}
Up to now, we have treated the dislocation as an infinitely long, perfectly straight line,
i.e., it travels from one end of the sample to the other ending on a surface defect ("open strings"). But more generally, we are interested in closed strings which reduce the symmetry.
Nonetheless, we may consider elliptical dislocations that mimic a finite length dislocation,
which we can take to be along the z-axis without loss of generality. The finiteness of the
string implies that momentum conservation in the $z-$direction is relaxed by an amount of order $1/L$, where $L$ is the length scale of the sample.

For scattering off a dislocation segment that runs from $z=0$ to $z = L$, the marginal operator that we have been working with becomes:
%\beq\int dt\int_0^L dz \psi^\dagger\psi(t, 0, 0, z)\eeq
%Upon Fourier transforming, the vertex is modified compared to the vertex derived in the idealized limit \eqref{clean_vertex}:
\beq\label{leaky_vertex}
S=\int dt \int [d^3q][d^3k] \hspace{2pt}g(\theta_0; \phi_k-\phi_q)\Big(\int_0^{L}dz e^{i(k_z-q_z)z}\Big)\psi^\dagger(t,\vec q)\psi(t,\vec k),
\eeq
where exact longitudinal momentum conservation is relaxed. 
It is straightforward to compute the factor in brackets, which is a sinc structure on the longitudinal momentum transfer. It admits an approximation as a Heaviside function, which we adopt to simplify subsequent calculations:
\beq \Big(\int_0^{L}dz e^{i(k_z-q_z) z}\Big)\sim L\Theta(\frac{\pi}{L}-|k_z-q_z|). \eeq
The normalization is chosen such that as $L\rightarrow \infty$, we recover $2\pi\delta(k_z-q_z)$, which enforces exact longitudinal momentum conservation.

We can perform the one-loop integral exactly, bypassing the need to break the EFT into multiple regions. The one-loop correction $\Sigma(\vec k_{in}=\vec k)$ to $g$ is now given by
\beq \Sigma(\vec k_{in}=\vec k)=\sum_n g_n^2 \frac{k_F}{2\pi v_F}\Big(\log\frac{|(\omega-v_F(k_{z}-k_F))^2-v^2_F\Lambda^2_z|}{v^2_F\Lambda^2})
\Big), \hspace{5pt}\Lambda_z  = \frac{\pi}{L}.\eeq
This result agrees with the infinite-dislocation result in the two regions Eqs. (\ref{one}) and (\ref{two}).
 In the equatorial region, where the particle–hole contributions are symmetric and the ``clean'' (momentum-conserving infinite length dislocation) theory produces no logarithm, introducing a finite leakiness scale does not generate one.
The finite length of the dislocation provides an additional infrared cutoff to the RG flow. A convenient way to organize the analysis is to treat the geometric proximity to the pole, $|k_F-|k_z||$, and the finite-length scale on equal footing. Both act as IR regulators that can terminate the logarithmic enhancement of the coupling.
Depending on the sign of $g_n$, each component may or may not grow in the IR. Whether the theory reaches the strong coupling scale  $T^{\star} \sim \Lambda e^{-g_n^\star}$
now depends on whether $T^\star > v_F \text{max}(1/L,\mid k_F-|k_z| \mid$).

\section{Conclusion}

In this paper, we have pointed out a new mechanism by which impurity scattering can lead to a non-trivial RG flow in analogy to the Kondo
impurity paradigm. What makes the present analysis distinct from the Kondo problem is that the way in which the particle-hole
non-cancellation is generated is not through non-commutativity, nor through changing the ground state to break the symmetry. Rather, the mechanism
is kinematic in nature and is due to the anisotropy generated by the codimension-two defect itself. 
The coupling is a function of the angle between the incoming momentum and the direction of the defect, and the running will be non-trivial for
$\theta_0\in \{0,\pi\}$. 

A finite length (elliptical) dislocation leads to 
momentum non-conservation in the longitudinal direction and acts as an IR cutoff on the RG time, as do deviations from $\theta_0=0$ or $\pi$.
For realistic materials, the typical loop size of a dislocation is orders of magnitude larger than $1/k_F$ (the UV cutoff of the EFT), so we expect many decades of running before
this scale is hit in a generic metal.

What is the nature of the IR fixed point? In the one-channel Kondo problem, the magnetic impurity is screened and leads to
a BCFT with a $\pi/2$ phase shift boundary condition, with $\pi/2$ being universal as a consequence of the Friedel sum rule, which relates the phase to
the particle number trapped by the defect. 
In the multi-channel Kondo,  there are dynamical degrees of freedom still active on the world-line and if they couple
in a relevant way to the electrons, one expects non-Fermi liquid behavior, which is what does/does not happen in the over/under screened cases \cite{Affleck:1990iv}.
Over-screening occurs when $k>2S$, $k$ being the number of channels and $S$ is the spin of the impurity.
 The distinction between the under-screened and screened cases will show up in the spin susceptibility.

 What does this teach us about the case at hand? We have argued that for $E<\omega^{d}_D$ the dislon coupling to electrons 
 is irrelevant and thus we expect that the low-energy fixed point will behave analogously to the screened Kondo problem, except that the  large scattering phase shift will not necessarily be $\pi/2$. One could imagine that the dislocation will form a bound
 state with a Fermi sea electron, but this will not change the qualitative behavior of the system unless the bound state
 energy happens to sit on the Fermi surface. However, unlike the Kondo case, it is not clear in the codimension-two case whether there will be a resistivity minimum since
the strong coupling is only relevant at $\theta_0\in \{0,\pi\}$, i.e., near forward scattering.  It would be nice to make a definitive statement
about this. However, we hesitate to do so since in the case of finite-size defects, momentum along the direction
of the dislocation is not conserved, so while the longitudinal resistivity will be suppressed by powers of $\delta k_z \sim 1/L$ the result will also
be enhanced by strong coupling. A clear answer to this requires a detailed transport calculation.
 
The most pertinent open question is: what happens when the dislon becomes relevant? As previously mentioned,
there is no a priori reason why the phonon and dislon Debye frequencies could not form a large ratio, in which case there
will be an RG window in which the phonons are still irrelevant but the dislon coupling to the electrons rapidly (power law)
grows strong as a consequence of the non-derivative nature (in scaling) of the electron-dislon coupling. 
If the fixed point
is hit at $E>\omega_D^{d}$, we would expect a non-trivial BCFT to be reached and result in non-Fermi liquid behavior as in the
overscreened Kondo problem.

\section{Acknowledgments}
The authors would like to thank Vlad Kozii, Shubhayu Chatterjee and Mike Widom for helpful discussions.
This work is supported by the Department of Energy (Grant No. DE- FG02-04ER41338 and FG02-06ER41449).

\end{document}